\documentclass[11pt,onecolumn,amssymb,nofootinbib]{revtex4}
\usepackage{amsmath, amsthm, amscd, amssymb}
\usepackage{bm}
\usepackage{bbm}

\begin{document}

\title{\bf Gravitation and the noise needed in objective reduction models}

\author{Stephen L. Adler}
\email{adler@ias.edu} \affiliation{Institute for Advanced Study,
Einstein Drive, Princeton, NJ 08540, USA.}

\begin{abstract}
I briefly recall intersections of my research interests with those of John Bell.
I then argue that the noise needed in theories of objective state vector reduction
most likely comes from a fluctuating complex part in the classical spacetime metric,
that is, state vector reduction is driven by {\it complex number valued}
``spacetime foam''.

\end{abstract}

\maketitle

\section{Introduction}

My research interests have intersected those of John Bell three times.  The first was when
I found the forward lepton theorem for high energy neutrino reactions, showing that
for forward lepton kinematics, a conserved vector current (CVC) and partially conserved
axial vector current (PCAC) imply that the neutrino cross section can be related to
a pion scattering cross section \cite{forwardlept}.  This led to an exchange of letters and discussions
with Bell during 1964-1965, which are described in the Commentaries for my selected
papers \cite{commentaries1}. The second was in the course of my work on the axial-vector
anomaly \cite{anomaly}, when I had further correspondence with John Bell, as described in
both the Commentaries \cite{commentaries2} and in the volume of essays on Yang-Mills
theories assembled by 't Hooft \cite{50years}.  The third time was a few years after Bell's death in
1990, when I became interested in the foundations of quantum theory and the quantum
measurement problem, in the course of writing my book on quaternionic quantum mechanics \cite{quaternion}.
Foundational issues in standard, complex quantum theory had preoccupied Bell for much of his career, and
led to his best known work.  However, my correspondence with Bell in the 1960s never touched on quantum foundations, and
I only read his seminal writings on the subject much later on.  In this article I focus on this third area of shared interests.

\section{Objective reduction models}

There is now a well-defined phenomenology of state vector reduction, pioneered by the work of Ghirardi, Rimini, and
Weber (GRW) and of Philip Pearle, and worked on by many others.  John Bell was interested in this program from the outset.
His 1987 essay ``Are there quantum jumps?'' \cite{bell} is devoted to a discussion of the GRW model, and Bell states ``For myself,
I see the GRW model as a very nice illustration of how quantum mechanics, to become rational, requires only a change
which is very small (on some measures!)''.

Current formulations of objective reduction models use not the discrete
localizations of the original GRW paper but rather  a nonlinear coupling of the Schr\"odinger equation to a stochastic noise
variable, as introduced in the continuous spontaneous localization (CSL) model of
Ghirardi, Pearle, and Rimini \cite{gpr}). The structure of this model, as I noted in my book on emergent quantum theory \cite{adlerbook}, is uniquely fixed by two
natural physical requirements, which might be expected to arise in an integral way from a more fundamental physical theory.  The first is
the requirement of state vector normalization  -- the unit norm of the state vector should be maintained in time.  The second is the
requirement that there should be no faster than light signaling -- the density matrix averaged over the noise should satisfy a linear
evolution equation of Lindblad form.  The form of the stochastic equation that is fixed by imposing these two requirements has
the satisfying feature that reduction to definite outcomes, with probabilities obeying the Born rule, can be proved.  This proof
was first given in \cite{gpr}, and was extended to include the L\"uders rule for degenerate systems in Adler, Brody, Brun and Hughston \cite{abbh}.

For the noise in the CSL model to give state vector reduction, as opposed to a unitary evolution, it must be introduced as an anti-Hermitian Hamiltonian
term which acts linearly on the wave function; norm preservation then requires the presence of a compensating quadratic term as well.  To achieve localization,
the noise is coupled to a local density operator.  There is then an important  constraint, since as shown by Pearle and Squires  \cite{psq} and
subsequent papers of Collett et al. \cite{collett} and Pearle et al. \cite{psq++}, the noise coupling must be mass-proportional to avoid conflicts with experiment.  This
suggests a noise coupling to the local mass density as the favored form of the CSL model.  Thus we have, in the extension of the CSL model to non-white noises \cite{adlerbassi},
\begin{equation}\label{eq:CSL}
\frac{d|\psi(t)\rangle}{dt}=\Big[ -iH+ \sqrt{\gamma} \int d^3x M(\vec x) \Phi(\vec x,t) + O\Big] |\psi(t)\rangle~~~,
\end{equation}
with $M(\vec x)$ the mass density for particles of masses and coordinates $m_i$ and $\vec q_i$,
\begin{equation}
M(\vec x)=\sum_i m_i \delta^3(\vec x-\vec q_i)~~~,
\end{equation}
and with $O$ denoting nonlinear terms that preserve state vector normalization.
Here $\Phi(\vec x, t)$ is a classical noise field, with expectation values ${\cal E}$ giving the mean and autocorrelation
\begin{equation}
{\cal E}[\Phi(\vec x,t)]=0~,~~{\cal E}[\Phi(\vec x,t_1) \Phi(\vec y,t_2)]=D(\vec x-\vec y,t_1-t_2)~~~,
\end{equation}
where $D(\vec x-\vec y,t_1-t_2)$ is the noise correlation function, and $\sqrt{\gamma}$ is a coupling constant
which could be absorbed into the definitions of $\Phi$ and of the correlation function $D(\vec x-\vec y,t_1-t_2)$.

\section{What is the physical origin of the noise?}

We now turn to the crucial question of what is the physical origin of the noise.  Some possible
cosmological particle physics origins of the noise were discussed in the second paper of \cite{adlerbassi}, but here
I want to broach another possibility, that the noise arises from a rapidly fluctuating complex
part of the classical gravitational metric $g_{\mu \nu}$.  (In scalar-tensor theories of gravitation,
the scalar field that accompanies the metric could also play a role.)  Let us suppose that the
classical metric has form
\begin{equation}\label{eq:Ansatz}
g_{\mu \nu}= \overline{g}_{\mu \nu} + \phi_{\mu \nu}~~~,
\end{equation}
with $\overline{g}_{\mu \nu}$ the conventional real space-time metric, and with the line element given as usual by
\begin{equation}
(ds)^2=\overline{g}_{\mu \nu} dx^{\mu}dx^{\nu}~~~.
\end{equation}
We assume that the extra part  $\phi_{\mu \nu}$ is an irreducibly {\it complex} fluctuation term,
with nonzero imaginary part, and
with expectations ${\cal E}$ given by
\begin{align}
{\cal E}[\phi_{\mu \nu}]=&0~~~,\cr
{\cal E}[\phi_{00}(\vec x, t_1) \phi_{00}^*(\vec y,t_2)]=&D(\vec x-\vec y,t_1-t_2)~~~,\cr
{\cal E}[\phi_{00}(\vec x, t_1) \phi_{00}(\vec y,t_2)]=&U(\vec x-\vec y,t_1-t_2)~~~.\cr
\end{align}
From the definition of the matter stress-energy tensor $T^{\mu \nu}$, the variation of the matter interaction  action $\delta S_{\rm int}$
produced by the fluctuating term in the metric is
\begin{equation}
\delta S_{\rm int} = - \frac{1}{2}\int d^4x (^{(4)}\overline{g})^{1/2} T^{\mu \nu} \phi_{\mu \nu}~~~,
\end{equation}
with $(^{(4)}\overline{g})^{1/2}$ the square root of the determinant of $-\overline{g}_{\mu \nu}$.
This action variation corresponds to minus one times the time integral of a variation in the matter interaction Hamiltonian of
\begin{equation}\label{eq:hint}
\delta H_{\rm int}=  \frac{1}{2} \int d^3x  (^{(4)}\overline{g})^{1/2} T^{\mu \nu} \phi_{\mu \nu}~~~.
\end{equation}
Since $T^{00}$ is proportional to the local mass density, in a flat spacetime with $\overline{g}_{\mu \nu}$
the Minkowski metric,
the coupling of the imaginary part of the $\phi_{00}$  term in the metric
gives a real noise coupling corresponding to $\sqrt{\gamma} \Phi$ of Eq.  \eqref{eq:CSL}.

In writing Eq. \eqref{eq:Ansatz} I am assuming that the both terms in the metric,
$\phi_{\mu \nu}$ as well as $ \overline{g}_{\mu \nu}$,  are symmetric in the indices $\mu,\, \nu$.
An alternative way of introducing a complex metric, the K\"ahler metric with Hermitian metric
$g_{\mu \nu}^* = g_{\nu \mu}$, has an imaginary part which is antisymmetric in its indices, and
so does not couple to the symmetric stress-energy tensor.
The Ansatz I that am making for the metric is similar to the definition
of so-called ``space-time foam'', except that usual ``foam'' fluctuations are
assumed to be real valued if classical, or self-adjoint if of quantum
origin. Instead, I am taking the fluctuation terms to be purely classical with
a non-zero imaginary part.

Before proceeding to further discussion of the Ansatz of Eq. \eqref{eq:Ansatz}, I note that it differs
substantively  from  previous proposals to relate state vector reduction to
gravitation, which have been reviewed by Shan Gao \cite{gao} in an article which critiques the well-known
proposal of Penrose and Di\'{o}si. (For a recent survey of the  Di\'{o}si-Penrose proposal and references see L.  Di\'{o}si \cite{diosi}).
Equation \eqref{eq:Ansatz}, augmented by requirements of state vector normalization and no faster than light signaling, gives
the usual de-correlation function  (in Di\'{o}si's term ``catness'' function) of the CSL
model. The Di\'{o}si-Penrose proposal, when incorporated into Di\'{o}si's universal position localization
model \cite{diosiuniversal}, gives a different de-correlation function which is related \cite{diosi} to the Newtonian gravitational potential.    Shan Gao's review \cite{gao}
also notes other articles suggesting a relation between state vector reduction and gravitation.  Pearle and Squires \cite{psqires}
propose to relate the CSL noise to fluctuations in the Newtonian potential or the curvature scalar, but do not explicitly address the
issue of hermiticity properties of the noise when viewed as an addition to the Hamiltonian.  K\'arolyh\'azy and subsequent collaborators \cite{karoly} try to relate wave function phase fluctuations induced by real-valued fluctuations
in the metric to the localizations needed in the GRW model.  However, within the framework of the CSL model, real-valued metric fluctuations lead to unitary
state vector evolution, and do not give state vector reduction.

\section{Arguments for a classical, but complex-valued metric}
Although much effort has been devoted to trying to quantize gravitation, there has been considerable discussion in
the literature of whether gravity has to be quantized.  Feynman \cite{feyn}, in his 1962-1963 lectures on gravitation,
notes the possibilities both that gravity may not have to be quantized, and that quantum theory may break down at large distances for macroscopic objects.
Dyson \cite{dyson} argues that the Bohr-Rosenfeld argument
for quantization of the electromagnetic field does not apply to gravity, and moreover, by a number of examples, shows that  it is hard
(perhaps not possible)  to formulate an experiment that can detect a graviton. A similar conclusion about detectability of gravitons
has been given by Rothman and Boughn \cite{rothman}.  Dyson also notes that arguments that have been cited to show that gravity
must be quantized really only show the inconsistency of a  particular model for classical gravity coupled to quantized matter, the M{\o}ller and Rosenfeld  semi-classical Einstein equation $G_{\mu \nu} = -8\pi G\langle \psi|T_{\mu \nu}|\psi \rangle$.  In a recent paper \cite{adlertrace}, I
argued that in trace dynamics pre-quantum theory, the metric should be introduced as a $c$-number in order for there to be an invariant volume element
defined through the determinant of the metric. (For a generalization of \cite{adlertrace} that allows a quantized metric
in trace dynamics, see Appendix A below.) In the trace dynamics framework, a consistent coupling of classical gravity to pre-quantum matter, that obeys the Bianchi and covariant conservation identities, is obtained by writing the classical Einstein equation as
\begin{equation}
G_{\mu \nu} = -\frac {8\pi G}{{\rm Tr}(1)} {\bf T}_{\mu \nu}~~~,
\end{equation}
with ${\bf T}_{\mu \nu}$ the covariantly conserved trace stress-energy tensor.

The possibility that the metric is complex valued has been considered previously in the literature; see \cite{munkhammar} for references.  The
formalism of general relativity, involving the construction of both the affine connection and the curvature tensor, is polynomial in the metric and its derivatives,
and does not impose a restriction that the metric be real-valued. In fact, one could argue that just as in the analysis of polynomial algebraic and differential equations, an
extension of the metric from the real number field to the complex number field is natural.
Since macroscopic bodies have a real stress-energy tensor, they serve as a source only for the real part of the Newtonian gravitational potential
$(g_{00}-1)/2$, and an imaginary part of the Newtonian gravitational potential would not be excited by them. Hence an imaginary part of the metric would not
change gravitational astrophysics.   An analysis that I carried out
with Ramazano\v glu \cite{static} of spherically symmetric, Schwarzschild-like solutions in trace dynamics-modified gravity, shows that in polar
coordinates the metric component $g_{00}$ develops a square root branch cut and becomes complex below a finite radius.  Although the branch cut turns
out to be a coordinate singularity, and $g_{00}$ is real-valued in isotropic coordinates, this calculation suggests that the metric $g_{\mu\nu}$
should be considered as a complex-valued classical field.

\section{Classical noise, not quantum noise, is needed for state vector reduction}

In this section I argue that quantum noise, unlike classical noise, does not lead to state vector reduction. I begin by contrasting the
kinematic structures of classical and quantum noise.

In the case of classical noise acting on a system $S$ in a
Hilbert space ${\cal H}_S$, the pure state density matrix $\rho_{\alpha}=|\psi_{\alpha}\rangle\langle \psi_{\alpha}|$, and its general
matrix element $\rho_{\alpha;ij}=\langle i|\rho_{\alpha}|j\rangle$, are functions of  a classical noise variable $\alpha$. One can then form an order $n$ density tensor,
with ${\cal E}$ the expectation over the noise variable,
\begin{equation}\label{eq:class}
\rho^{(n)}_{i_1j_1,i_2j_2,...,i_nj_n}={\cal E}[\rho_{\alpha;i_1j_1}\rho_{\alpha;i_2j_2}....\rho_{\alpha;i_nj_n}]~~~
\end{equation}
which satisfies various identities derived by contracting pairs of indices \cite{hier}.  The hierarchy of all such tensors captures the properties of the noise acting on the
system;  in particular, the expectation ${\cal E}[V]$ of the variance $V={\rm Tr}\rho A^2 -({\rm Tr}\rho A)^2$ of an operator $A$ can be rewritten as
\begin{equation}\label{eq:classvar}
{\cal E}[V]={\rm Tr}(\rho^{(1)} A^2)-\rho_{i_1j_1,i_2j_2}^{(2)}A_{j_1i_1}A_{j_2i_2}~~~.
\end{equation}
The proof of state vector reduction of \cite{gpr} and \cite{abbh} can be recast in terms of the Lindblad evolution
satisfied by the density matrix $\rho^{(1)}=\rho$, and special properties of the density tensor $\rho^{(2)}$ for the nonlinear evolutions that obey
the conditions of norm preservation and no faster than light signaling.

To define quantum noise, we consider a closed system, consisting of a system $S$ and an environment $E$ in an overall pure state, characterized by
a density matrix $\rho$.  Quantum noise
consists of the environmental fluctuations acting on $S$ that are averaged over when the environment is traced over.  Labeling system states by $|i\rangle,\,|j\rangle$
and environmental states by $|e_a\rangle,\, a=1,2,...$, we can define a density tensor $\rho^{(n)}$ by
\begin{align}\label{eq:quant}
\rho^{(n)}_{i_1j_1,i_2j_2,...,i_nj_n}=&{\rm Tr}_E\rho_{i_1j_1}\rho_{i_2j_2}...\rho_{i_nj_n}~~~,\cr
{\rm Tr}_E {\cal O}=&\sum_a \langle e_a | {\cal O}|e_a \rangle~~~,\cr
\end{align}
with $\rho_{i_{\ell}j_{\ell}}$ the matrix acting on the environmental Hilbert space according to
\begin{equation}
(\rho_{i_{\ell}j_{\ell}})_{e_qe_r}=\langle e_qi_{\ell}|\rho|e_rj_{\ell}\rangle  ~~~.
\end{equation}
This quantum hierarchy does not obey all of the properties of the classical hierarchy of Eq. \eqref{eq:class}. In particular, the variance $V$ defined
by $V={\rm Tr}( \rho A^2)- ({\rm Tr} \rho A)^2$, with ${\rm Tr}$ the trace over the full system plus environment Hilbert space, cannot be put in a form
analogous to Eq. \eqref{eq:classvar}, and one cannot construct a proof of reduction following the method used in the classical case.  This should not be
a surprise, since Bassi and Ghirardi have proved \cite{bassighirardi},  using only linearity of the Schr\"odinger evolution, that quantum evolution by itself cannot
give rise to state vector reduction.

Can this conclusion be evaded by decreeing that the system plus environment form an open system, which is not in a pure state?  I do not believe so,
when the finite propagation of signals is taken into account \cite{response}.  Consider a system, consisting of a Stern-Gerlach molecular beam apparatus,  plus its environment,
enclosed in a large container, which is at a distance from from the apparatus much greater than $3 \times 10^7$cm, and which is used to conduct the following thought experiment.
Let the container consist of perfectly reflecting boundaries, which can be simultaneous opened by command from synchronized timers just inside.  When the boundaries are closed,
the interior is in a pure state and the non-reduction proof of \cite{bassighirardi} applies.  When the boundaries are open, the interior forms an open
system since photons and other particles (generically, information) can get out.  The typical time for a molecular beam experiment is $10^{-3}$s, which is
more than the time for the Stern-Gerlach apparatus inside to be informed whether the boundaries are open or closed.  So I do not see how the state of the boundaries,
and thus whether the system is closed or open,
can influence the outcome of the Stern-Gerlach experiment being conducted inside.

\section{Conclusion}

I have argued that there is a natural confluence between the noise requirements of objective theories
of state vector reduction, and the noise that can be furnished by a fluctuating, classical, complex valued
component of the gravitational metric.

\section{Acknowledgements}

Completion of this work was supported in part by the National Science Foundation under Grant No. PHYS-1066293 and the hospitality of the
Aspen Center for Physics.  I wish to thank Shan Gao for conceiving the volume in which this essay appears, and for his editorial assistance.

\appendix

\section{Generalization to a quantized metric}

To extend the argument of \cite{adlertrace} to a quantized metric, we note that under
a general coordinate transformation from coordinates $x_{\mu}$ to $x^{\prime}_{\mu}$ the metric changes
according to
\begin{equation}\label{eq:metrictrans}
g^{\prime}_{\mu \nu}= g_{\rho\sigma}\frac{\partial x^{\rho}}{\partial x^{\prime \mu}} \frac{\partial x^{\sigma}}{\partial x^{\prime \nu}} ~~~.
\end{equation}
Since in trace dynamics the coordinates $x_{\mu}$ and $x^{\prime}_{\mu}$ are $c$-numbers, when the metrics $g^{\prime}_{\mu \nu}$ and $g_{\rho\sigma}$
are matrix-valued Eq. \eqref{eq:metrictrans} applies separately to their traces and their trace-free parts.  That is, writing
\begin{align}
g^{\prime}_{\mu \nu}=&g^{\prime}_{{\rm CLASSICAL}\,\mu \nu}+g^{\prime}_{{\rm PRE-QUANTUM}\, \mu \nu} ~~~,\cr
g^{\prime}_{{\rm CLASSICAL}\,\mu \nu}=&\frac{{\rm Tr}(g^{\prime}_{\mu \nu})}{{\rm Tr}( 1)}~~~,\cr
{\rm Tr}(g^{\prime}_{{\rm PRE-QUANTUM}\, \mu \nu}) =&\,0~~~,\cr
\end{align}
and
\begin{align}
g_{\rho \sigma}=&g_{{\rm CLASSICAL}\,\rho \sigma}+g_{{\rm PRE-QUANTUM}\, \rho \sigma} ~~~,\cr
g_{{\rm CLASSICAL}\,\rho \sigma}=&\frac{{\rm Tr}(g_{\rho \sigma})}{{\rm Tr}( 1)}~~~,\cr
{\rm Tr}(g_{{\rm PRE-QUANTUM}\, \rho \sigma}) =&\,0~~~,\cr
\end{align}
we have the transformation laws
\begin{align}\label{eq:metrictans}
g^{\prime}_{{\rm CLASSICAL}\,\mu \nu}=& g_{{\rm CLASSICAL}\,\rho\sigma}\frac{\partial x^{\rho}}{\partial x^{\prime \mu}} \frac{\partial x^{\sigma}}{\partial x^{\prime \nu}} ~~~, \cr
g^{\prime}_{{\rm PRE-QUANTUM}\, \mu \nu}=& g_{{\rm PRE-QUANTUM}\, \rho\sigma}\frac{\partial x^{\rho}}{\partial x^{\prime \mu}} \frac{\partial x^{\sigma}}{\partial x^{\prime \nu}} ~~~.\cr
\end{align}
We see that the trace part of the matrix-valued metric $g_{\mu \nu}$ , which we have defined here as $g_{{\rm CLASSICAL}\,\mu \nu}$, has the same transformation law
as the full matrix-valued metric, and so the determinant of the trace part can be used to define an invariant volume element.  The classical metric $g_{\mu\nu}$ of the text of
this article, and of \cite{adlertrace}, can thus be identified with the trace part $g_{{\rm CLASSICAL}\,\mu \nu}$ of the matrix-valued metric.

The arguments given in \cite{adlertrace}, for the form of the effective action functional of the classical
metric arising from pre-quantum fluctuations, require that the action for the pre-quantum fields be Weyl scaling invariant.
When a trace-free pre-quantum metric $g_{{\rm PRE-QUANTUM}\, \mu \nu}$ is included, Weyl scaling of the metric takes the form
\begin{align}
g_{{\rm CLASSICAL}\,\mu\nu}(x) &\to \lambda^2 g_{{\rm CLASSICAL}\,\mu \nu}(x)~~~,\cr
g_{\rm CLASSICAL}^{\mu\nu}(x) &\to \lambda^{-2} g_{\rm CLASSICAL}^{\mu \nu}(x)~~~,\cr
g_{{\rm PRE-QUANTUM}\,\mu\nu}(x) &\to \lambda^2 g_{{\rm PRE-QUANTUM}\,\mu \nu}(x)~~~,\cr
g_{\rm PRE-QUANTUM}^{\mu\nu}(x) &\to \lambda^{-2} g_{\rm PRE-QUANTUM}^{\mu \nu}(x)~~~, \cr
\end{align}
where indices on both the classical and the pre-quantum metric are raised using the classical metric,
\begin{align}
g_{{\rm CLASSICAL}}^{ \rho\sigma}=&g^{\rho \gamma}_{\rm CLASSICAL} g^{\sigma \delta}_{\rm CLASSICAL} g^{}_{{\rm CLASSICAL}\,\gamma\delta}~~~,\cr
g_{{\rm PRE-QUANTUM}}^{ \rho\sigma}=&g^{\rho \gamma}_{\rm CLASSICAL} g^{\sigma \delta}_{\rm CLASSICAL} g^{}_{{\rm PRE-QUANTUM}\,\gamma\delta}~~~.\cr
\end{align}
The requirement of Weyl scaling invariance of the pre-quantum gravitational action can now be satisfied by fourth order trace actions such as
\begin{align}
{\bf S}_{\rm GRAV~PRE-QUANTUM}=& \int d^4x \big(-{\rm Det}(g_{{\rm CLASSICAL}\,\xi \eta})\big)^{1/2} g_{\rm CLASSICAL}^{\alpha \beta} g_{\rm CLASSICAL}^{\gamma \delta}\cr
&\times {\rm Tr} \left(\frac{\partial g_{{\rm PRE-QUANTUM}\, \rho\sigma}}{\partial x^{\alpha} \partial x^{\gamma}} \frac{\partial g_{{\rm PRE-QUANTUM}}^{ \rho\sigma}}{\partial x^{\beta} \partial x^{\delta}}\right)~~~.\cr
\end{align}
The classical part of the metric is taken, as in \cite{adlertrace}, to have a standard Einstein-Hilbert action,
\begin{equation}
S_{\rm GRAV~CLASSICAL}=\frac{1}{16\pi G} \int d^4x \big(-{\rm Det}(g_{{\rm CLASSICAL}\,\xi \eta})\big)^{1/2} R[ g_{{\rm CLASSICAL}\,\xi \eta}]~~~.
\end{equation}
Because the gravitational constant $G$ is dimensional, this action  is not Weyl scaling invariant.

\end{document}